\documentclass[usenatbib]{aa}
\usepackage[varg]{txfonts}

\usepackage[T1]{fontenc}
\usepackage{ae,aecompl}

\def\figdir{./Hill-figs/}

\usepackage{graphicx}	
\usepackage{amsmath}	
\usepackage{amssymb}	
\usepackage{mathtools}
\usepackage{array}
\usepackage{color}
\usepackage{bm}
\usepackage{mathrsfs}
\usepackage{dcolumn}

\renewcommand{\epsilon}{\varepsilon}                                                        
\def\norm#1{\left\Vert#1\right\Vert}

\def\Frac#1#2{{{\displaystyle\strut#1}\over{\displaystyle\strut#2}}}

\def\tr{{\bf\tilde r}}

\def\r{{\bf r}}

\def \rot {r_{12}}

\def\bea{\begin{eqnarray}}
\def\eea{\end{eqnarray}}
\def\be{\begin{equation}}
\def\ee{\end{equation}}
\def\pdix#1{\times10^{#1}}
\def \dix#1{10^{#1}}

\def\nnb{\nonumber\\}

\def\O{\mathrm{O}}

\usepackage{nicefrac}

\def\H{\mathcal{H}}

\def\Gr{\mathcal{G}}
\def \G {{\bf G}}

\def\rC{{\mathscr{C}}}
\def \Cc{C_c}
\def \Ch{C_c^{\mathrm H}}

\def \Cmmr{C_c^{\mathrm{MMR}}}

\def \P {\mathcal{P}}

\AtBeginDocument{\mathcode`v=\varv}

\title{Hill stability in the AMD framework}

\author{
Antoine C. Petit, Jacques Laskar, Gwena\"el Bou\'e
}

\titlerunning{Hill stability in the AMD framework}
\authorrunning{Petit}

\institute{
	IMCCE, CNRS-UMR8028, Observatoire de Paris, PSL University, Sorbonne Université, 77 Avenue Denfert-Rochereau, 75014 Paris,
	France\\
	\email{antoine.petit@obspm.fr}
}

\date{Accepted XXX. Received YYY; in original form ZZZ}

\abstract{In a two-planet system, due to Sundman (1912) inequality, a topological boundary can forbid close encounters between the two planets for infinite time. A system is said Hill stable if it verifies this topological condition.
	Hill stability is widely used in the study of extra solar planets dynamics. However people often use the coplanar and circular orbits approximation.\\
	In this paper, we explain how the Hill stability can be understood in the framework of Angular Momentum Deficit (AMD).
	In the secular approximation, the AMD allows to discriminate between a priori stable systems and systems for which a more in depth dynamical analysis is required.
	We show that the general  Hill stability criterion can be expressed as a function of only the semi major axes, the masses and the total AMD of the system.
	The proposed criterion is only expanded in the planets-to-star mass ratio $\epsilon$ and not in the semi-major axis ratio, in eccentricities nor in the mutual inclination. Moreover the expansion in $\epsilon$ remains excellent up to values of about $10^{-3}$ even for two planets with very different mass values.\\
	We performed numerical simulations in order to highlight the sharp change of behaviour between Hill stable and Hill unstable systems. We show that Hill stable systems tend to be very regular whereas Hill unstable ones often lead to rapid planet collisions. We also remind that Hill stability does not protect from the ejection of the outer planet.
}

\begin{document}

\maketitle

\section{Introduction}

The chaotic nature of planetary dynamics has made necessary the development of stability criteria because the complete study of individual systems for the lifetime of the central star would require too much computational power.
Moreover, in the context of exoplanets dynamics, the orbital parameters are often not known with great precision, making impossible to conduct a precise dynamical study.
For multi-planetary systems, the best solutions yet are empirical stability criteria based on minimum spacing between planets obtained from numerical simulations \citep{Chambers1996,Pu2015}.
For tightly packed systems, \cite{Tamayo2016} has suggested a solution based on short integrations and a machine-learning algorithm.

Another approach consists in the study of the angular momentum deficit \citep[AMD, see][]{Laskar1997,Laskar2000}.
The AMD is a weighted sum of the planets' eccentricities and mutual inclinations and can be interpreted as a dynamical temperature of the planetary system.
The AMD is conserved at all order of averaging over the mean motions. In \citep{Laskar2017}, it was shown that if the AMD is small enough, collisions are impossible.
We can define a sufficient stability condition from the secular conservation of the AMD, the AMD-stability.
AMD-stable systems are long-lived whereas a more in depth dynamical study is necessary for AMD-unstable systems.

The initial AMD-stability definition is based on the secular approximation. In \citep{Petit2017}, the criterion was  slightly modified in order to exclude systems experiencing short term chaos due to first order mean motion resonances (MMR) overlap.
The MMR overlap is the main source of chaos and instability in planetary dynamics. 
Based on \cite{Chirikov1979} resonance overlap criterion, \cite{Wisdom1980} 
has derived an analytical stability criterion for the two-planet systems that since, have been widely used.
Wisdom's criterion is obtained for circular and coplanar planets. 
It was improved to take into account moderate eccentricities \citep{Mustill2012,Deck2013}. We show in \citep{Petit2017} that this MMR
overlap criterion  can be expressed solely as a function of the AMD, semi-major axes and masses for almost coplanar and low eccentricity systems.

The MMR overlap criterion gives a clear limit between regular and chaotic orbits for small eccentricities.
However, this criterion is based on first order expansions in the planets-to-star mass ratio, the spacing between the planets, eccentricities and inclinations. 
As a result, for wider orbit separations, the secular collision criterion from \citep{Laskar2017} remains a better limit \citep{Petit2017}. 
Moreover, the MMR overlap criterion only takes into account the interaction between a couple of planets, making it really accurate only for two-planet systems.

In the case of two-planet systems, the topology of the phase space gives a far simpler criterion of stability, the Hill stability.
Based on a work by \cite{Sundman1912} on the moment of inertia in the three-body problem, \cite{Marchal1975} note the existence of forbidden zones in the configuration space.
In \citep{Marchal1982}, they extend the notion of Hill stability to the general three-body problem and show that some systems can forbid close encounters between the outer body and any of the inner bodies.
In particular, the Hill stability ensures that collisions  between the outer body and the close binary (in a two-planet system, between the outer planet and the inner planet or the star) are impossible for infinite time.

The results of \citep{Marchal1975,Marchal1982} have many applications outside of the Hill stability.
One can cite a sufficient condition for the ejection of a body from the system \citep{Marchal1984,Marchal1984a} or the determination of the limit of triple close approach for bounded orbits \citep{Laskar1984}.

\cite{Marchal1982}  present the planetary problem (one body with a much larger mass) as a particular case, but the result was mainly popularized by \cite{Gladman1993} who introduced a minimal spacing for initially circular and coplanar systems.
Gladman's paper also proposes some criteria for eccentric orbits in some particular configurations of masses.
Gladman's result was refined in order to cover other situations as, for example, the case of inclined orbits \citep{Veras2004}.
See \citep{Georgakarakos2008} for a review of stability criteria for hierarchical three-body problems.

The Hill stability is consistent with numerical integrations, where a sharp transition between Hill stable and Hill unstable systems is often observed \citep{Gladman1993,Barnes2006,Deck2013}.
Moreover, \cite{Deck2013} analyze the differences between the MMR overlap and Hill stability criteria.
 They remark that there exists an area where orbits are chaotic due to the overlap of MMR but long lived due to Hill stability.
 
In this paper, we show that \citet{Marchal1982} Hill stability criterion fits extremely well in the AMD-stability framework.
In section \ref{sec.theory}, we derive a criterion for Hill stability solely expressed as a function of the total AMD, the semi-major axes and the masses.
Our criterion does not need any expansion in the spacing, eccentricities or inclinations of the orbits and admits all previous Hill criteria as particular approximations.
In order to do so we follow the reasoning proposed by \cite{Marchal1982}.

We then compare the Hill stability criterion with the AMD-stability criteria proposed in \citep{Laskar2017,Petit2017}.
In the last section, we carry numerical integrations of two-planet systems over a large part of the phase space.
We show that the only parameters of importance are the initial AMD and semi-major axis ratio.

\section{Hill stability in the 3-body problem}
\label{sec.theory}

\subsection{Generalised Hill curves}
Let us use the formalism proposed in \citep{Marchal1975,Marchal1982} for the definition of the Hill regions for the general 3-body problem.
We mainly consider the planetary case and therefore adapt slightly their notations to this particular problem.
Let us consider two planets of masses $m_1, m_2$ orbiting a star of mass $m_0$.
Let $\Gr$ be the constant of gravitation, $\mu=\Gr m_0$,
\begin{equation}
\epsilon=\frac{m_1+m_2}{m_0},
\label{eps}
\end{equation}
the planet mass to star mass ratio and 
\begin{equation}
	\gamma=\frac{m_1}{m_2}.
\end{equation}
As in \cite{Laskar2017}, we use the heliocentric canonical coordinates $(\tr_j,\r_j)_{j=1,2}$. In those coordinates, the Hamiltonian is
\begin{align}
\H=\sum_{j=1}^{2}\left(\frac{1}{2}\frac{\norm{\tr_j}^2}{m_j}-\frac{\mu m_j}{r_j}\right)+
  \frac{1}{2}\frac{\norm{\tr_1+\tr_2}^2}{m_0}-\frac{\Gr m_1m_2}{\rot},
  \label{ham}
\end{align}
where $\rot =\|\r_1-\r_2\|$.
We denote $\G$ the total angular momentum of the system
\begin{equation}
\G=\sum_{j=1}^{2}\r_j\times\tr_j,
\end{equation}
and $G$ its norm. From now, we assume $\G$ to be aligned along the $z$-axis.
We also define
\begin{equation}
	\Lambda_j=m_j\sqrt{\mu a_j},
\end{equation}
where $a_j$ is the semi-major axis of the $j$th planet. We also note $e_j$, the eccentricity and $i_j$, the inclination of the orbital plane with the horizontal plane.
The AMD, $C$ \citep{Laskar1997,Laskar2000} has for expression
\begin{equation}
C=\sum_{j=1}^{2} \Lambda_j\left(1-\sqrt{1-e_j^2}\cos(i_j)\right).
\end{equation}

Following \citep{Marchal1982}, we define a generalised semi-major axis
\begin{equation}
	a=-\frac{\Gr M^*}{2\H},
	\label{def.a}
\end{equation}
and a generalised semi-latus rectum
\begin{equation}
p=\frac{m_0(1+\epsilon)}{\Gr M^{*2}}G^2,
\label{def.p}
\end{equation}
where
\begin{equation}
M^*=m_0m_1+m_0m_2+m_1m_2=m_0^2\epsilon\left(1+\epsilon\frac{\gamma}{(1+\gamma)^2}\right).
\end{equation}
$a$ and $p$ are the two length units that can be built from the first integrals $\H$ and $G$.

We finally define two variable lengths. First, $\rho$ the mean quadratic distance
\begin{equation}
M^* \rho^2=m_0m_1r_1^2+m_0m_2r_2^2+m_1m_2\rot^2,
\end{equation}
that is proportional to the moment of inertia $I$ computed in the centre of mass frame \citep{Marchal1982}
\begin{equation}
	I=\frac{1}{2}\frac{M^*\rho^2}{m_0+m_1+m_2}.
\end{equation}
We also define $\nu$, the mean harmonic distance
\begin{equation}
\frac{M^*}{\nu}=\frac{m_0m_1}{r_1}+\frac{m_0m_2}{r_2}+\frac{m_1m_2}{\rot},
\end{equation}
which is proportional to the potential energy
\begin{equation}
U=-\frac{\Gr M^*}{\nu}.
\end{equation}

For a system with given $\H$ and $G$, some configurations of the planets are forbidden. Indeed, the value of the ratio $\rho/\nu$ is constrained by the inequality \citep{Marchal1975}
\begin{equation}
	\frac{\rho}{\nu}\geq \frac{\rho}{2a}+\frac{p}{2\rho},
	\label{ine.MaSa}
\end{equation}
derived from Sundman's inequality \citep{Sundman1912}.
Moreover, if the system has a negative energy, the right hand side of (eq. \ref{ine.MaSa}) has a minimum value obtained for ${\rho=\sqrt{ap}}$. We therefore have the inequality \citep{Marchal1982}
\begin{equation}
	\frac{\rho^2}{\nu^2} \geq \frac{p}{a}= -\frac{2m_0(1+\epsilon)}{\Gr^2 M^{*3}} \H G^2.
	\label{ine.MaBo}
\end{equation}
If $p/a$ is high enough, the inequality (\ref{ine.MaBo}) makes some regions of the phase space inaccessible.
In this case, we can ensure that certain initial conditions forbid collisions between the two planets for all times.

\begin{figure}[tbp]
	\includegraphics[width=9cm]{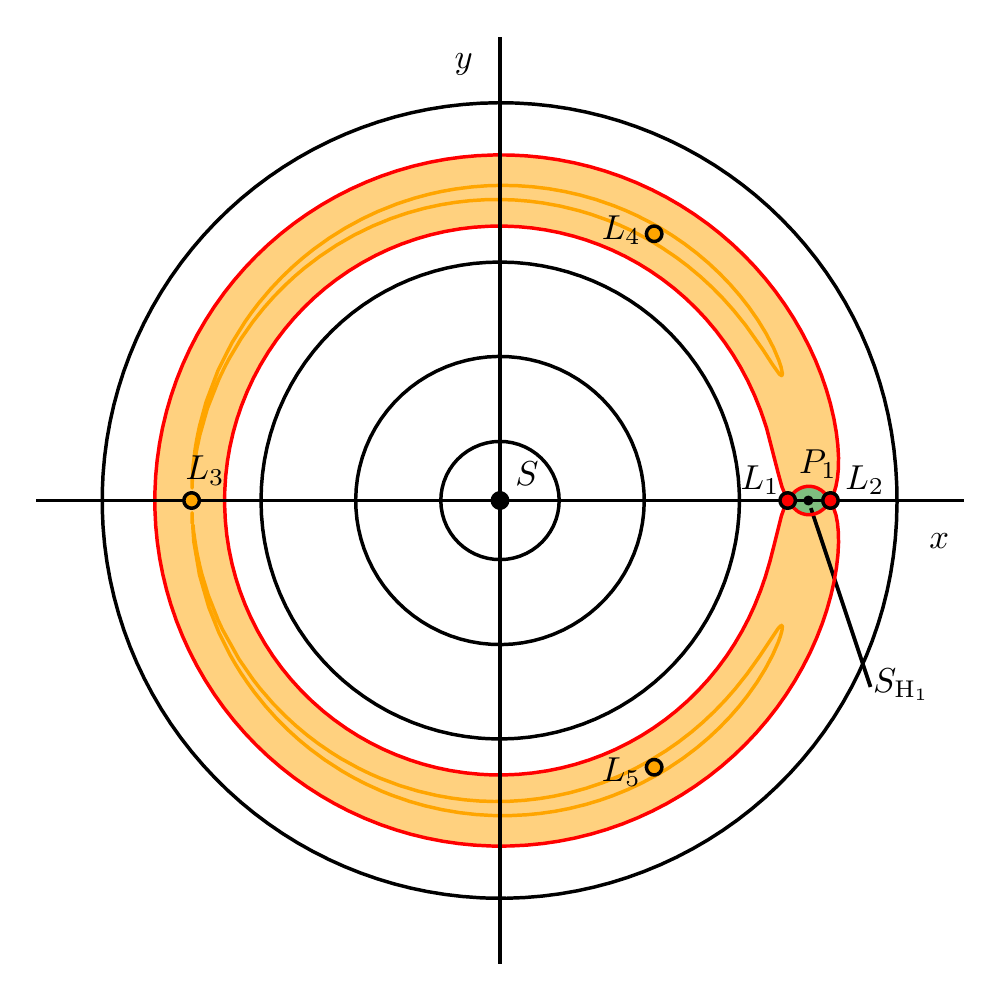}
	\caption{We represent in the $\P$ plane, some levels of the function $R$ defined in (\ref{funR}) for $\epsilon=\dix{-3}$ and $\gamma=1$.
	The two red points correspond to the Lagrange points $L_1$ and $L_2$ and the three orange points to $L_3$, $L_4$ and $L_5$. The orange-filled area corresponds to the region where $R(x,y)<R(L_1)$.
	The star $S$ is at the origin, the planet $P_1$ at (1,0) and (x,y) are the coordinates of the second planet.
	The green region $S_\mathrm{H_1}$ represents the Hill sphere of the first planet.
}
	\label{figrsn}
\end{figure}

Let us study the values of the function $(\rho/\nu)^2$. Since this ratio only depends on the ratios of mutual distances, we can always place ourselves in the plane generated by the three bodies.
We can also choose to place the first planet on the $x$-axis and normalize the lengths by $r_1$.
Let us call this plane $\P$ and note $(x,y)$ the coordinates of the second planet.
In the plane $\P$ (see Figure~\ref{figrsn}), the star $S$ is at the origin, the first planet $P_1$ is situated at the point $(1,0)$.
For the planar restricted 3-body problem, this reduction is equivalent to study the dynamics in the corotating frame.
We note
\begin{equation}
R(x,y)=\left(\frac{\rho}{\nu}\right)^2.
\label{funR}
\end{equation}

The shape of the function $R$ in  $\P$ only depends on the mass distribution, e.g., the two ratios $\epsilon$ and~$\gamma$.
In Figure~\ref{figrsn}, we can see level curves of the function $R$ plotted in $\P$ for $\epsilon=10^{-3}$ and $\gamma=1$.
$R$ is minimal and equal to 1 at the two Lagrange points $L_4$ and $L_5$ and goes to $+\infty$ for $\|\r_2\|\to 0$, $\|\r_2-\r_1\|\to 0$ or $\|\r_2\|\to+\infty$.
The function has 3 saddle nodes at the Lagrange points $L_1,L_2$ and $L_3$. We give the method to compute the position of the Lagrange points in appendix \ref{app.R}. Let us denote respectively $x_1,x_2$ and $x_3$ their abscissa.
At the lowest order in $\epsilon$ we have
\begin{align}
x_1 &= 1- \left(\frac{\epsilon}{3}\right)^{1/3}+\O(\epsilon^{2/3}),\nnb
x_2 &=  1 + \left(\frac{\epsilon}{3}\right)^{1/3}+\O(\epsilon^{2/3}), \label{Lag.abscissa}\\
x_3 &= - 1+\frac{7}{12}\frac{\gamma-1}{\gamma+1}\epsilon +\O(\epsilon^2).\nonumber
\end{align}

At the lowest order in $\epsilon$, the value of $R$ at the Lagrange points are
\begin{align}
	R(L_1)&=1+3^{4/3}\epsilon^{2/3}\gamma/(\gamma+1)^2+\O(\epsilon)\nnb
	R(L_2)&=1+3^{4/3}\epsilon^{2/3}\gamma/(\gamma+1)^2+\O(\epsilon),\\
	R(L_3)&=1+2\epsilon\gamma/(\gamma+1)^2 +\O(\epsilon^2)\nonumber.
\end{align}
$R(L_1)$ and $R(L_2)$ have the same first order term but differ in the expansion of higher order.
Indeed, if  $m_0\geq m_1\geq m_2$, we have $R(L_1)\geq R(L_2)$ \citep{Marchal1982}.
From now, let us assume ${R(L_1)\geq R(L_2)}$ (if not, we can just substitute $R(L_2)$ to $R(L_1)$ in further equations).

For $p/a\geq R(L_1)$, the accessible domain is split in three parts, the Hill sphere of the star $S$ which is around the origin, the first planet Hill sphere\footnote{The Hill region is usually called the Hill sphere although it is not technically a sphere.} $S_\mathrm{H_1}$ (in green on Figure \ref{figrsn}) and the outer region.
In this case, if the second planet is not initially inside $S_\mathrm{H_1}$, it will never be able to enter this region.
However, if $P_2$ is in the outer region, the Hill stability cannot constrain the possibility of ejection.
Similarly, if $P_2$ is closer to the star (in the inner region), a collision with the star is still possible.

The study of the function $R$ and the inequality (\ref{ine.MaBo}) gives a non-collision criterion for an infinite time. Marchal and Bozis called it the Hill stability.

\begin{proposition}[\citealp{Marchal1982}\label{prop.MaBo}]
	Let us consider a negative energy three-body problem with a body $S$ of mass $m_0$ and two others $P_1$ and $P_2$ of mass $m_1$ and $m_2$ such that $m_0\geq m_1\geq m_2$.
	We place ourselves in the $\P$ plane defined by $S$, $P_1$ and $P_2$ (Figure \ref{figrsn}). If $P_2$ is not initially inside the Hill sphere $S_\mathrm{H_1}$ of $P_1$, the system is Hill stable if
	\begin{equation}
	\frac{p}{a}>R(L_1),
	\label{hilldef}
	\end{equation}
	where $a$ is defined in (\ref{def.a}), $p$ in (\ref{def.p}) and $R$ in (\ref{funR}).
\end{proposition}

From this inequality, \cite{Gladman1993} obtained criteria for initially circular orbits and for two particular cases of eccentric orbits: the case of equal masses and small eccentricities and the case of equal masses and large but equal eccentricities.

While Gladman's Hill stability criterion for initially circular orbits is useful, the eccentric criteria are too particular to be used in the context of a generic system.
It is however possible to obtain a very general Hill stability criterion using the AMD to take into account the eccentricities and inclinations of the orbits.

\subsection{AMD condition for Hill stability}

The total energy of the system can be written
\begin{equation}
	\H=-\frac{m_2^3\mu^2}{2\Lambda_2^2}\left(\frac{\gamma}{\alpha}+1+h_1\right),
	\label{energy}
\end{equation}
where $\alpha=a_1/a_2$ and
\begin{equation}
	h_1=-\frac{2\Lambda_2^2}{m_2^3\mu^2}\left(\frac{1}{2}\frac{\norm{\tr_1+\tr_2}^2}{m_0}-\frac{\Gr m_1m_2}{\rot}\right).
	\label{h1}
\end{equation}
From now on, we assume that initially $\alpha\leq1$ (if not we can just renumber the two planets).
Similarly the angular momentum can be rewritten
\begin{equation}
	G=\Lambda_2\left(\gamma\sqrt{\alpha}+1-\rC\right),
	\label{angmom}
\end{equation}
where 
\begin{equation}
\rC=\frac{C}{\Lambda_2}=\gamma\sqrt{\alpha}\left(1-\sqrt{1-e_1^2}\cos i_1\right)+1-\sqrt{1-e_2^2}\cos i_2
    \label{def.rC}
\end{equation}
 is the relative AMD \citep{Laskar2017}.
Combining (\ref{ine.MaBo}), (\ref{energy}) and (\ref{angmom}), we obtain
\begin{equation}
\frac{p}{a}=\frac{(1+\epsilon)}{(\gamma+1)^3(1+\epsilon\gamma/(\gamma+1)^2)^3}\left(\frac{\gamma}{\alpha}+1+h_1\right)\left(\gamma\sqrt{\alpha}+1-\rC\right)^2.
\end{equation}

The Hill stability criterion (\ref{hilldef}) can be rewritten without any approximation as a condition on $\rC$ and we have the following formulation of the Hill stability.
\begin{proposition}[Hill stability]
With the hypotheses of the proposition \ref{prop.MaBo}, assuming the elliptical elements can be defined (\emph{i.e.} both Keplerian energies are negative), a system is Hill stable if
 \begin{equation}
\rC<C_c^\mathrm{Ex}=\gamma\sqrt{\alpha}+1-(\gamma+1)^{3/2}\sqrt{\frac{ R(L_1)(1+\epsilon\gamma/(\gamma+1)^2)^3}{(1+\epsilon)\left(\gamma/\alpha+1+h_1\right)}},
\label{hillAMD}
\end{equation}
where $\rC$ is the relative AMD (\ref{def.rC}) and $h_1$ the normalized perturbation part (\ref{h1}).
\end{proposition}

The inequality (\ref{hillAMD}) is equivalent to the proposition \ref{prop.MaBo} but we isolated on the left hand side the contribution from the AMD.
Up to the perturbation term $h_1$, the right hand side of (\ref{hillAMD}) only depends on the masses and the semi-major axis ratio $\alpha$.
If we only keep the terms of leading order in $\epsilon$ in the square root of the right hand side of (eq.~\ref{hillAMD}), we obtain an expression that depends only on $\alpha$, $\epsilon$ and $\gamma$.

\begin{proposition}[Hill stability, planetary case\label{prop.HS}]
	For small enough $\epsilon$, a two-planet system is Hill stable if the  relative AMD $\rC$ verifies the inequality
	\begin{equation}
	\rC<\gamma\sqrt{\alpha}+1-(1+\gamma)^{3/2}\sqrt{\frac{\alpha}{\gamma+\alpha} \left(1+\frac{3^{4/3}\epsilon^{2/3}\gamma}{(1+\gamma)^2}\right)} +\O(\epsilon).
	\label{hillepsAMD}
	\end{equation}
\end{proposition}

As explained in appendix \ref{app.pertterm}, $h_1$ is of smaller order in $\epsilon$ and can be neglected if the criterion is verified.
We want to stress out that the expression (\ref{hillepsAMD}) is obtained with only an expansion in $\epsilon$ and does only depends of $\alpha$, $\rC$ and the masses of the bodies. 
The term of order $\epsilon^{2/3}$ in (eq. \ref{hillepsAMD}) also depends on $\gamma/(\gamma+1)^2$, but we show in appendix \ref{app.pertterm} that (\ref{hillepsAMD}) is still valid for $\gamma\ll1$ or $\gamma\gg1$.

\subsection{Close planets approximation}

Assuming $1-\alpha\ll 1$, and a small AMD value, further approximations can be made.
At leading order in $\rC,1-\alpha$ and $\epsilon$, the inequality (\ref{hillepsAMD}) becomes
\begin{equation}
\rC<\frac{3\gamma}{8(\gamma+1)}(1-\alpha)^2-\frac{3^{4/3}\gamma}{2(\gamma+1)}\epsilon^{2/3}.
\label{hillepsalp}
\end{equation}
One can isolate $1-\alpha$ in this expression to obtain an approximate minimum spacing for Hill stable systems.
\begin{proposition}[Hill~stability, close planets case]
	For a close planets system, the minimum spacing criterion for Hill stability is
	\begin{equation}
	\frac{a_2-a_1}{a_2}=1-\alpha>\sqrt{4\times3^{1/3}\epsilon^{2/3}+\frac{8}{3}\frac{\gamma+1}{\gamma}\rC}.
	\label{betterGladman}
	\end{equation}
\end{proposition}

Gladman's eccentric criteria can be recovered from (\ref{betterGladman}) if
the AMD is developed under the assumptions (same planet masses, small or large
and equal eccentricities) made in \citep{Gladman1993}. However (\ref{betterGladman}) is
more general as it takes into account mutual inclinations or uneven mass distribution.

In the case of circular orbits, we also get the well-known formula \citep{Gladman1993}
\begin{equation}
1-\alpha>2\times3^{1/6}\epsilon^{1/3}=2.40\epsilon^{1/3}.
\label{Gl.circ}
\end{equation}

\subsection{Comparison of the Hill criteria}

\begin{figure}[tbp]
	\includegraphics[width=9cm]{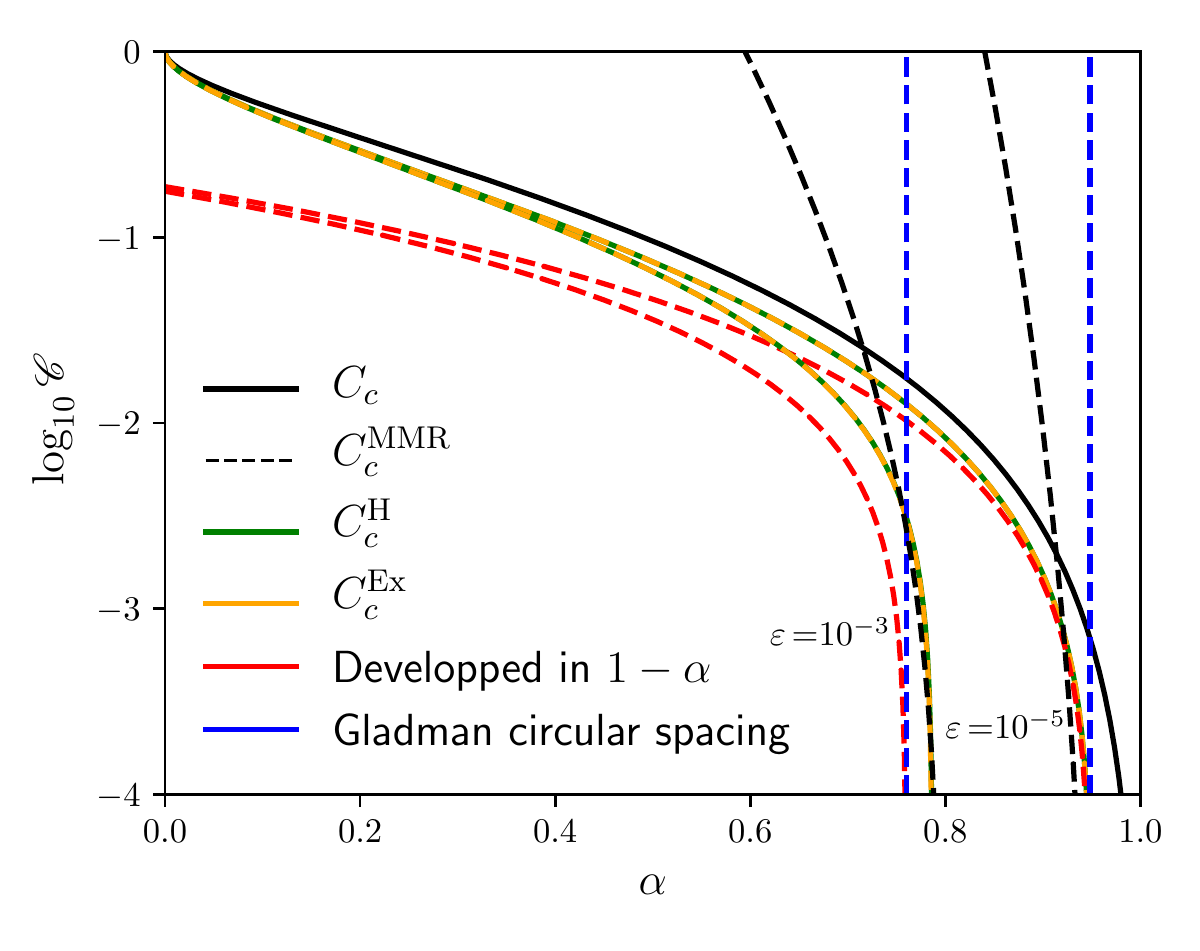}
	\caption{Comparison of the right hand sides of the inequalities (\ref{hillAMD}, orange), (\ref{hillepsAMD}, green) and (\ref{hillepsalp}, red) as a function of $\alpha$ for $\gamma=1$ and $\epsilon=\dix{-5}$ (upper three curves) or $\epsilon= \dix{-3}$ (lower three curves). The black curve is the critical AMD for the collision condition \citep{Laskar2017}. The green and orange curves are on top of eachother (see Figure \ref{critdiff}).
	The critical AMD from MMR overlap $\Cmmr$ and Gladman's (1993) circular criterion are also plotted for comparison.}
	\label{crithill}
\end{figure}

\begin{figure}[tbp]
	\includegraphics[width=9cm]{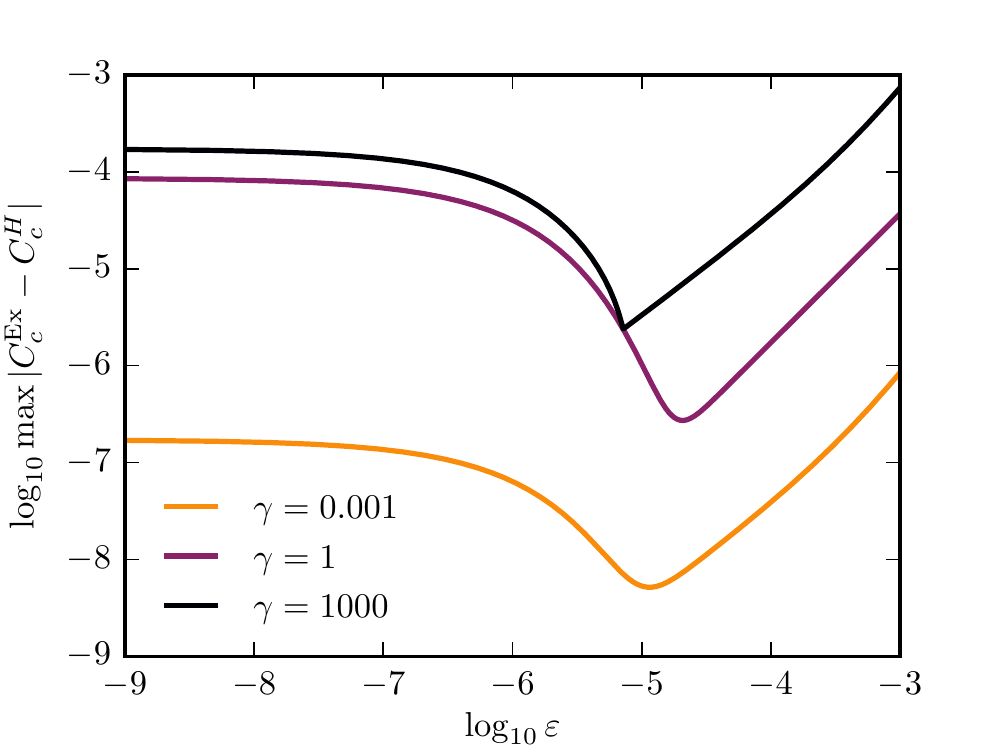}
	\caption{Maximum difference between $C_c^{\text{Ex}}$ (\ref{hillAMD}) and $\Ch$ (\ref{critAMDH}), respectively the right hand sides of (\ref{hillAMD}) and (\ref{hillepsAMD}), as a function of $\epsilon$ for various values of $\gamma=m_1/m_2$. In (\ref{hillAMD}), the term $h_1$ is evaluated using the approximation of the kinetic term given in appendix \ref{app.pertterm} }
	\label{critdiff}
\end{figure}

We can compare the right-hand side of (\ref{hillAMD}), (\ref{hillepsAMD}), (\ref{hillepsalp}) 
and Gladman's circular approximation (\ref{Gl.circ}) to test how relevant are the approximations made here.
In Figure \ref{crithill}, we plot the exact expression $C_c^{\mathrm{Ex}}$ (eq. \ref{hillAMD}, in green),
the expansion in $\epsilon$ (eq. \ref{hillepsAMD}, in orange), the approximation for close planets~(\ref{hillepsalp}, in red) as well as the minimum spacing for circular orbits (\ref{Gl.circ}, in blue).
We see that the expansion in $\epsilon$ (\ref{hillepsAMD}) cannot be distinguished from the exact curve (\ref{hillAMD}). In order to better quantify this, we plot in Figure \ref{critdiff} the maximum difference between the two curves as a function of $\epsilon$ for various values of the mass ratio $\gamma$.

We see in figure \ref{critdiff} that for the range of $\epsilon$ used in planetary dynamics 
(typically from $\dix{-6}$ to $\dix{-3}$), the expression (\ref{hillepsAMD}) developed in $\epsilon$ is accurate even for very uneven planet mass distribution. From now, we use (\ref{hillepsAMD}) to define the Hill stability.


\section{Comparison with the AMD-stability}
\label{sec.AMDcomp}
In \citep{Laskar2017}, the AMD-stability of a system  is defined by comparing its AMD to a critical value $\Cc$ above which the stability of the system cannot be guaranteed.
The AMD-stability is a sufficient condition for long-lived stability of a system.
The result from \citep{Laskar2017} was obtained by finding the minimal AMD needed for a system to allow for collisions in the secular approximation, \emph{i.e.}, assuming the semi major axes are fixed.
We completed in \citep{Petit2017} the AMD-stability definition by excluding the configurations such that the overlap of first order MMR occurs.
The modified critical AMD is hereafter noted $\Cmmr$.

We can use (\ref{hillepsAMD}) to define a critical AMD $\Ch$ for the Hill stability
\begin{equation}
\Ch=\gamma\sqrt{\alpha}+1-(1+\gamma)^{3/2}\sqrt{\frac{\alpha}{\gamma+\alpha} \left(1+\frac{3^{4/3}\epsilon^{2/3}\gamma}{(1+\gamma)^2}\right)}\ .
\label{critAMDH}
\end{equation}
A system will be Hill stable if its initial relative AMD is smaller than the initial critical AMD $\Ch$.

We see that for two planets, the Hill stability definition fits extremely well in the AMD-stability framework.
We can also compare $\Ch$ to the previously proposed critical AMD.
The collision critical AMD $\Cc$ is plotted in Figure \ref{crithill} with two values of $\Ch$ for $\epsilon=\dix{-5}$ (resp. $\dix{-3}$).
We can see that the Hill stability criterion is stricter ($\Ch<\Cc$) than the  collision condition for secular dynamics. 
It can be easily understood since the Hill stability forbids the planets to approach each-other.

Indeed, let us consider a Hill stable system, \emph{i.e.} such that ${\rC<\Ch}$. 
As a result the two planets cannot approach each other by less than their mutual Hill radius for any variation of semi major axes and thus also in
the secular system. In particular, a configuration such that the two orbits intersect is impossible.
Therefore, the system is AMD-stable and $\rC<\Cc$.
Since we are not making any additional hypothesis on $\rC$, we have $\Ch<\Cc$. The strict inequality comes from the positive minimal distance between the two planets.

As a comparison, we also plot in Figure \ref{crithill} the MMR critical AMD $\Cmmr$.
For small relative AMD $\rC$, $\Cmmr$ and $\Ch$ are almost identical. 

\section{Numerical simulations}

The Hill criterion proposed by \citet{Marchal1982} has already been tested numerically in particular cases \citep{Gladman1993,Veras2004,Barnes2006}.
In their comparison between the Hill and the overlap of MMR criteria, \cite{Deck2013} note  a sharp transition in the proportion of chaotic orbits at the Hill limit $(p/a)=(p/a)|_c$.
It also appears that for small $\epsilon$ and $\alpha$ close to 1, the overlap of MMR criterion provides a better limit for the chaotic region.

We want to test if the Hill criterion gives a good limit to the chaotic region for wider separations.
Moreover, we want our initial conditions to sample homogeneously the phase space.
Indeed, the Hill stability criterion studied in this paper only depends on few quantities,
the relative AMD $\rC$ and the ratio of semi-major axis and not the angles or the actual distribution of the AMD between the degrees of freedom of eccentricities or inclinations. 
Choosing an homogeneous sampling of the initial conditions also avoids giving too much importance to regions protected by MMR due to particular combinations of angles while another choice of angles would have given an unstable orbit.

\subsection{Numerical set-up}

We run numerical simulations using the symplectic scheme ABAH1064 from \citep{Farres2013}. We choose our initial conditions such that:
\begin{itemize}
	\item the outer planet semi-major axis $a_2$ is fixed at 1 au,
	\item the AMD and inner planet semi-major axis $a_1$ are chosen such that we have a regular grid in the plane $(\alpha,\sqrt{\rC})$. Such a scaling in $\rC$ is chosen to have an approximately uniform distribution in terms of eccentricities and inclination,
	\item the AMD is on average equipartitioned between the eccentricity and inclination degrees of freedom,
	\item the inclinations are chosen such that the angular momentum is on the $z$-axis,
	\item the angles are chosen randomly,
	\item the star mass is taken as $1 \mathrm{M}_\odot$ and the planets masses do not vary for each grid of initial conditions.
\end{itemize}
We then integrate each initial condition for 500 kyr using a time-step of $\dix{-3}$ yr.
The numerical integration is stopped if the planets approach each other by less than a quarter of their mutual Hill radius, if a planet reaches $\dix{-2}$~AU or 20~AU or if the relative variation of energy is higher than $\dix{-8}$.

In order to measure the stability of a system, we use the frequency map  analysis \citep{Laskar1990b,Laskar1993}. Our criterion is based on the relative variation of the main frequencies in the quasiperiodic best fit.
More precisely, let be $n_k^{(i)}$ (respectively $n_k^{(f)}$) be the frequency obtained by frequency 
analysis for the planet $k$ for the first (resp. last) 100 kyr of integration.
We consider an orbit to be chaotic if
\begin{equation}
\delta n =\max_k \left|\frac{n_k^{(f)}-n_k^{(i)}}{n_k^{(i)}}\right|
\label{deltan}
\end{equation}
is greater than $\dix{-4}$.
The chosen threshold is such that the variation of semi-major axis
changes by about 1\% in a few Gyr if we assume a constant diffusion process as it would happen for a random walk.
Indeed, $\delta n$ measures the variation of frequency over 500 kyr. If the diffusion rate remains constant, a variation of the order of 1\% will on average needs a time 10,000 times  larger, \emph{i.e.}, 5 Gyr.

If the integration time is shorter due to a collision or ejection, we set $\delta n$ to 1.
Since we draw randomly most of the initial parameters, we bin the results into a two-dimension grid in $(\alpha,\sqrt{\rC})$ and average the frequency variation in each bin.

\subsection{Results}

We first integrate 100,000 initial conditions on a uniform grid with $\alpha$ taking values from 0.5 to 1 and $\rC$ from 0 to 0.1.
The masses of the two planets are equal to $0.5\pdix{-5} \mathrm{M}_\odot$, so that $\epsilon=\dix{-5}$ and $\gamma=1$.
In this simulation, 78.4\% of the orbits survive up to 500 kyr, 21.1\% end up in a collision between the two planets and 0.5\% of the integrations are stopped because of the non-conservation of energy due to an unresolved close encounter.

\begin{figure}[tbp]
	\includegraphics[width=10cm]{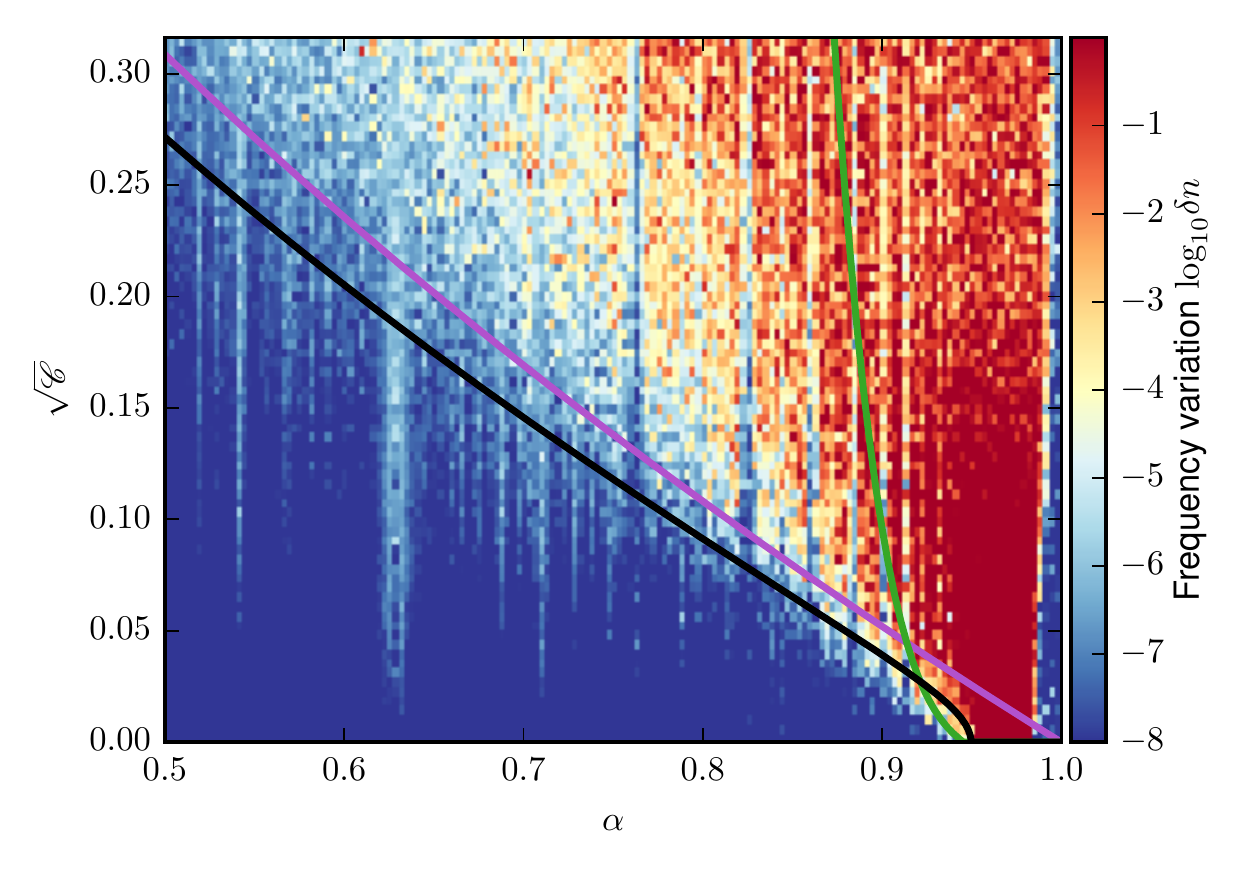}
	\caption{Frequency variation $\delta n$ (\ref{deltan}) for 100,000 initial conditions binned in a $160\times75$ grid with $\epsilon=\dix{-5}$ and $\gamma=1$.
		The black curve is the Hill critical AMD $\Ch$  (\ref{critAMDH}), the purple curve is the collisional critical AMD $\Cc$ \citep{Laskar2017} and the green one is the critical AMD obtained from the overlap of first order MMR $\Cmmr$ \citep{Petit2017}. Each bin is average over about 8 initial conditions.}
	\label{naf_5_1}
\end{figure}

\begin{figure}[tbp]
	\includegraphics[width=10cm]{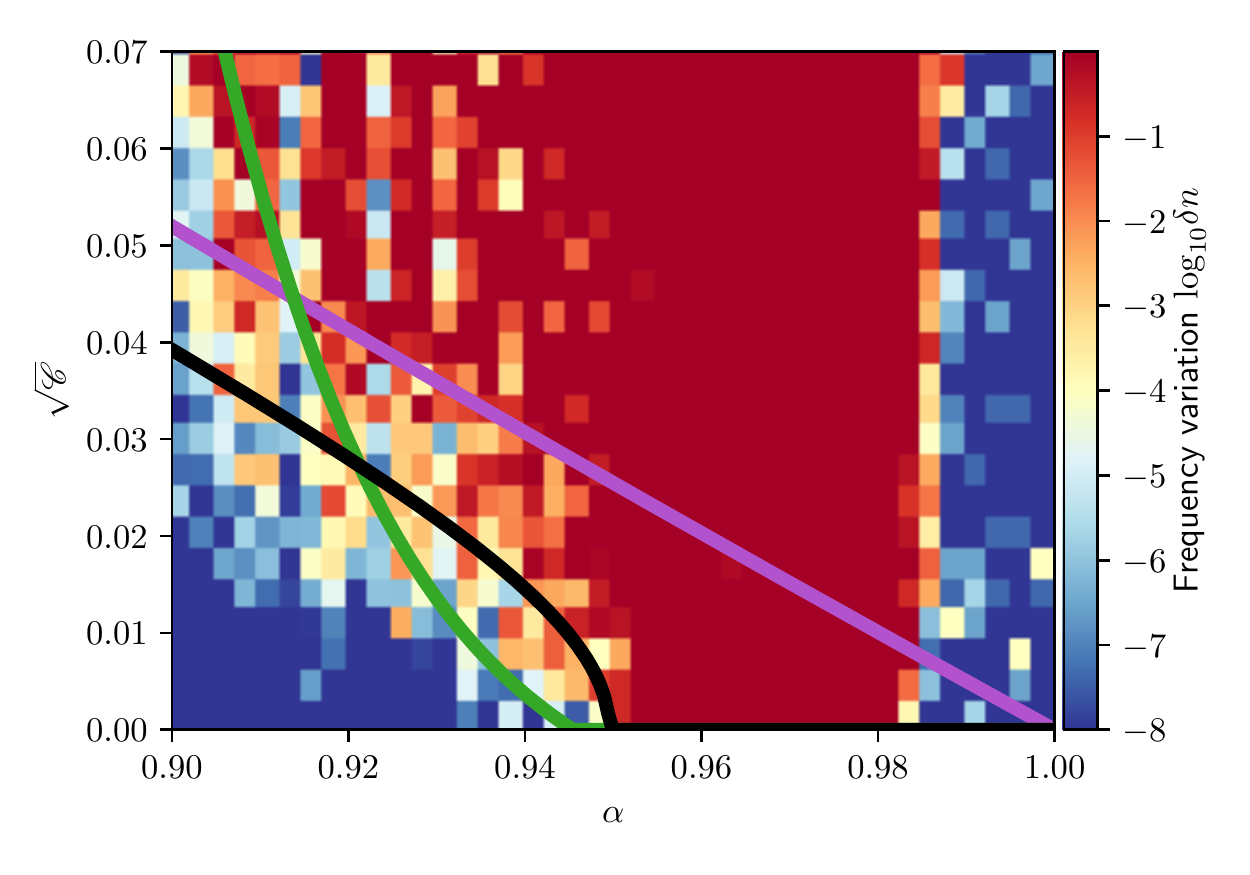}
	\caption{Zoom of Figure \ref{naf_5_1} for $0.9<\alpha<1$ and $\sqrt{\rC}<0.07$. In the region where $\Cmmr<\rC<\Ch$, we see that orbits are chaotic but Hill stable.}
	\label{naf_5_1_zoom}
\end{figure}

The results of the frequency analysis are shown in Figure~\ref{naf_5_1}. We see that the chaotic region is well constrained by the Hill curve $\Ch$. Indeed, very few orbits with $\rC<\Ch$ appear to be chaotic.
The region where Hill stable orbits ($\rC<\Ch$) are chaotic seems restricted to the region where ${\Cmmr<\rC<\Ch}$ around $\alpha \simeq 0.94$ and low $\rC$ \emph{i.e.}, for orbits experiencing MMR overlap (a zoom of Fig. \ref{naf_5_1} is given in Fig. \ref{naf_5_1_zoom} to see it more easily).
The behavior of planets initially in this region was already discussed in \citep{Deck2013}.
Orbits that are Hill unstable (${\rC>\Ch}$) appear to be largely chaotic up to some resonant islands situated at $\alpha\simeq1$ (co-orbital resonance) and near the 3:2 and 4:3 resonances ($\alpha \simeq 0.76$ and 0.82).
We also see that for larger separations (${\alpha\lesssim 0.6}$), orbits are less chaotic. However it is probable that for longer integration times, these orbits would end up unstable.

\begin{figure}[tbp]
	\includegraphics[width=9cm]{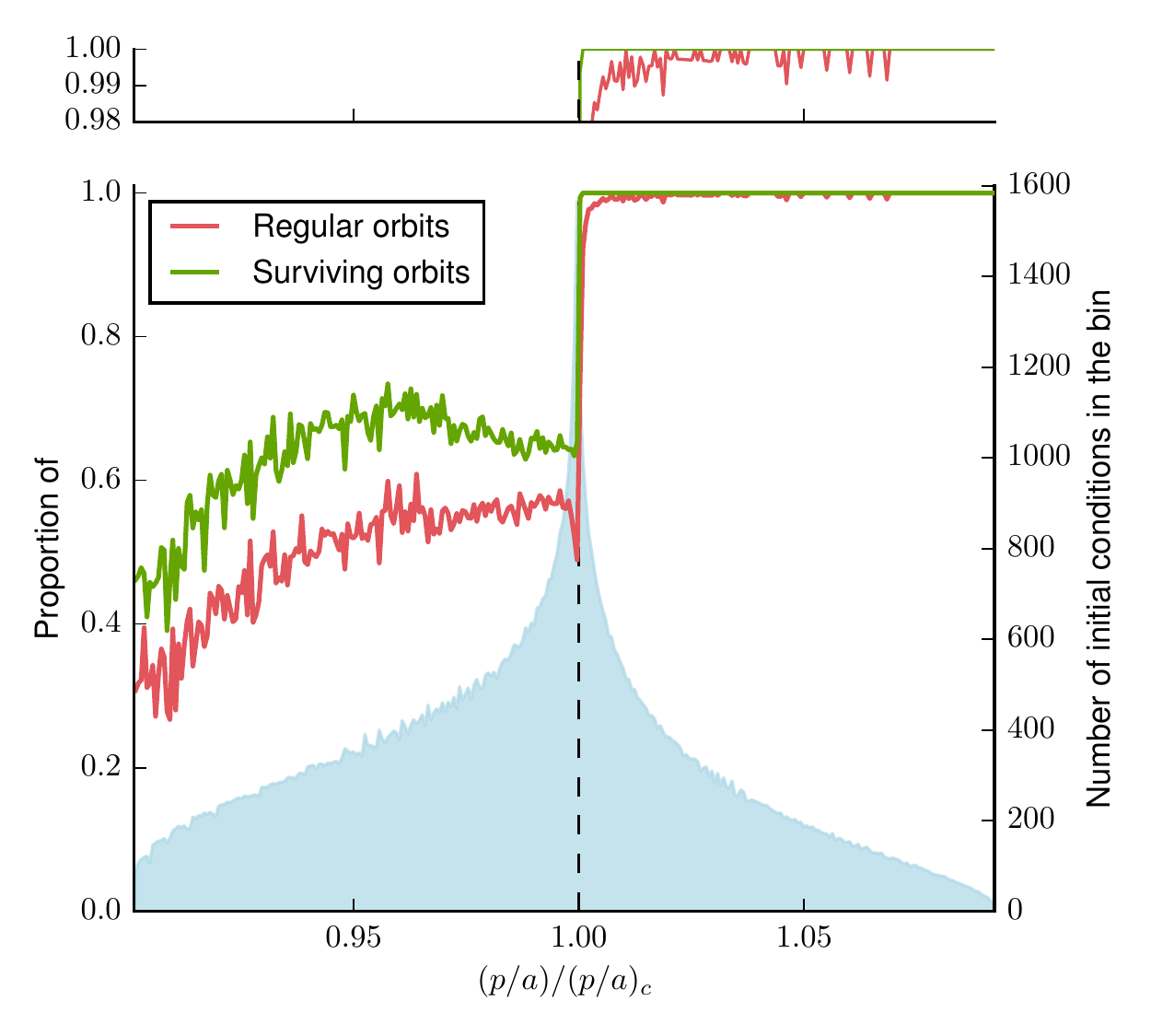}
	\caption{Proportion of orbits that survived for 500 kyr (green curve) and non chaotic orbits ($\delta n <\dix{-4}$, red curve) as a function of $(p/a)/(p/a)_c$.  $(p/a)/(p/a)_c>1$ means that the initial condition verifies the Hill criterion. Above of the main figure, we added a zoom of the upper part of the plot. The light blue histogram represents the number of initial conditions in each of the 300 bins used to compute the fractions plotted.} 
	\label{chaos_5_1}
\end{figure}

In order to highlight how the Hill criterion separates the chaotic orbits from the stable ones, we can 
plot the fraction of regular orbits as a function of $(p/a)/(p/a)_c$ as suggested in \citep{Barnes2006}.
We see in Figure~\ref{chaos_5_1} that there is a sharp limit at $p/a=(p/a)_c$. All Hill stable  integrations go to 500 kyr and very few are chaotic.
For Hill unstable orbits (${\rC>\Ch}$), we see a slight increase of regular and surviving orbits with $p/a$ but the decrease is not as significant as the change at the Hill limit.
On average 64.9\% of the Hill unstable orbits survive and 52.3\% are regular.

\begin{figure}[tbp]
	\includegraphics[width=9.5cm]{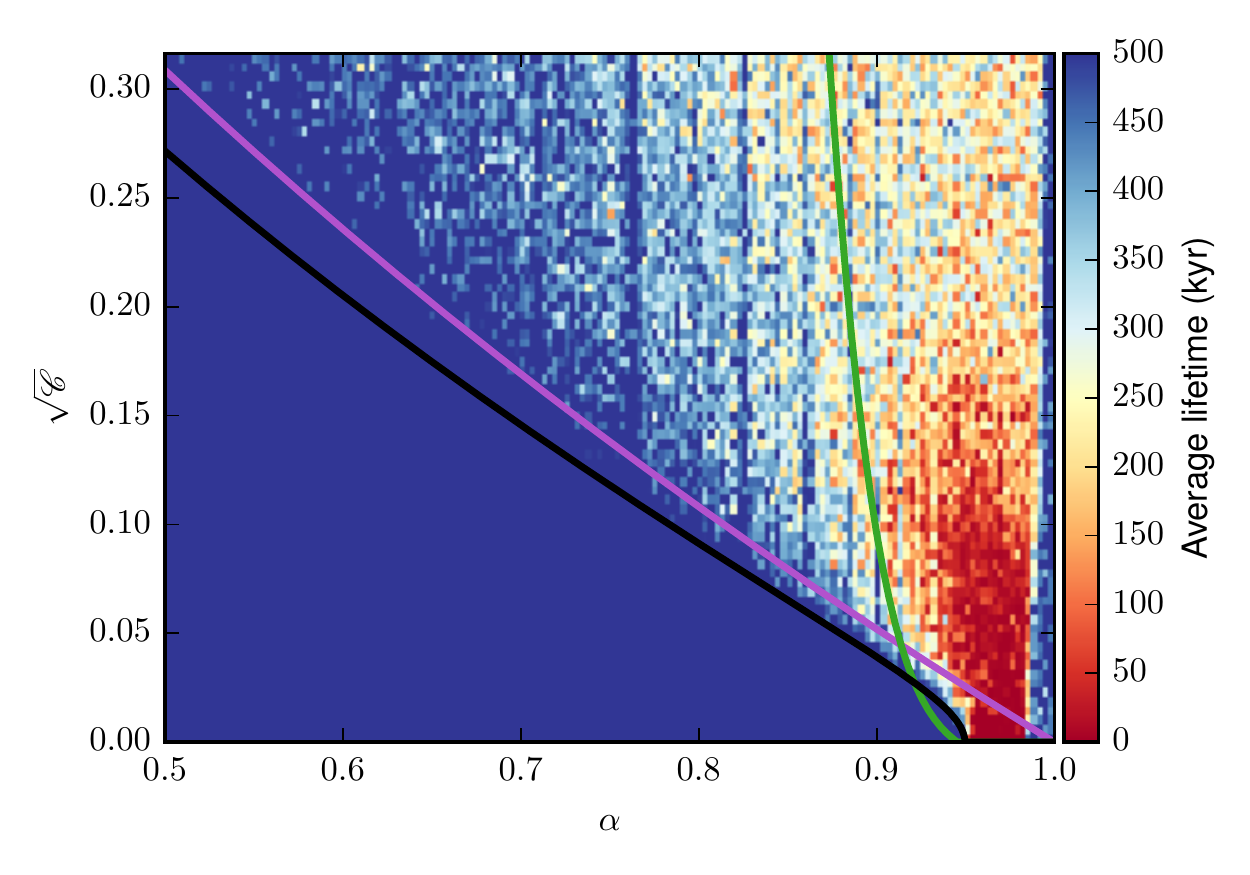}
	\caption{Average lifetime of the system as a function of $\alpha$ and $\rC$. The parameters are similar to figure \ref{naf_5_1}. Each bin represents an average over 8 initial conditions. A dark blue color implies that all integration ended at 500 kyr.} 
	\label{crash_5_1}
\end{figure}

We plot in Figure \ref{crash_5_1} the average time of the integrations as a function of initial $\alpha$ and $\rC$. The initial conditions are binned in a 160 $\times$ 75 grid.
We see in this figure that the average lifetime of a system is almost always smaller than 500 kyr in the Hill unstable region.
We also remark that the average lifetime is less than 50 kyr for a system with $\alpha\gtrsim 0.9 $ and not too large $\rC$ but increases for wider separations or greater AMD.
Indeed for higher $\rC$, the planets are on average initially further away from crossing orbits because the initial choice of eccentricities and inclinations is random.

\subsection{Influence of the masses}

To highlight the role of the mass of the planets, we run another 
simulation with the same method but with ${\epsilon =\dix{-3}}$ and $\gamma=1$.
We integrate 40,000 initial conditions and do a similar analysis.
After 500 kyr, 54.6\% of the systems lead to a collision between the two planets, 2\% to an ejection and 0.5\% are stopped due to non conservation of energy.
If we consider only the Hill unstable systems (${\rC>\Ch}$), only 24.0\% have survived and 15.7\% are regular.
As expected, larger masses lead to much more unstable systems.

\begin{figure}[tbp]
	\includegraphics[width=9.5cm]{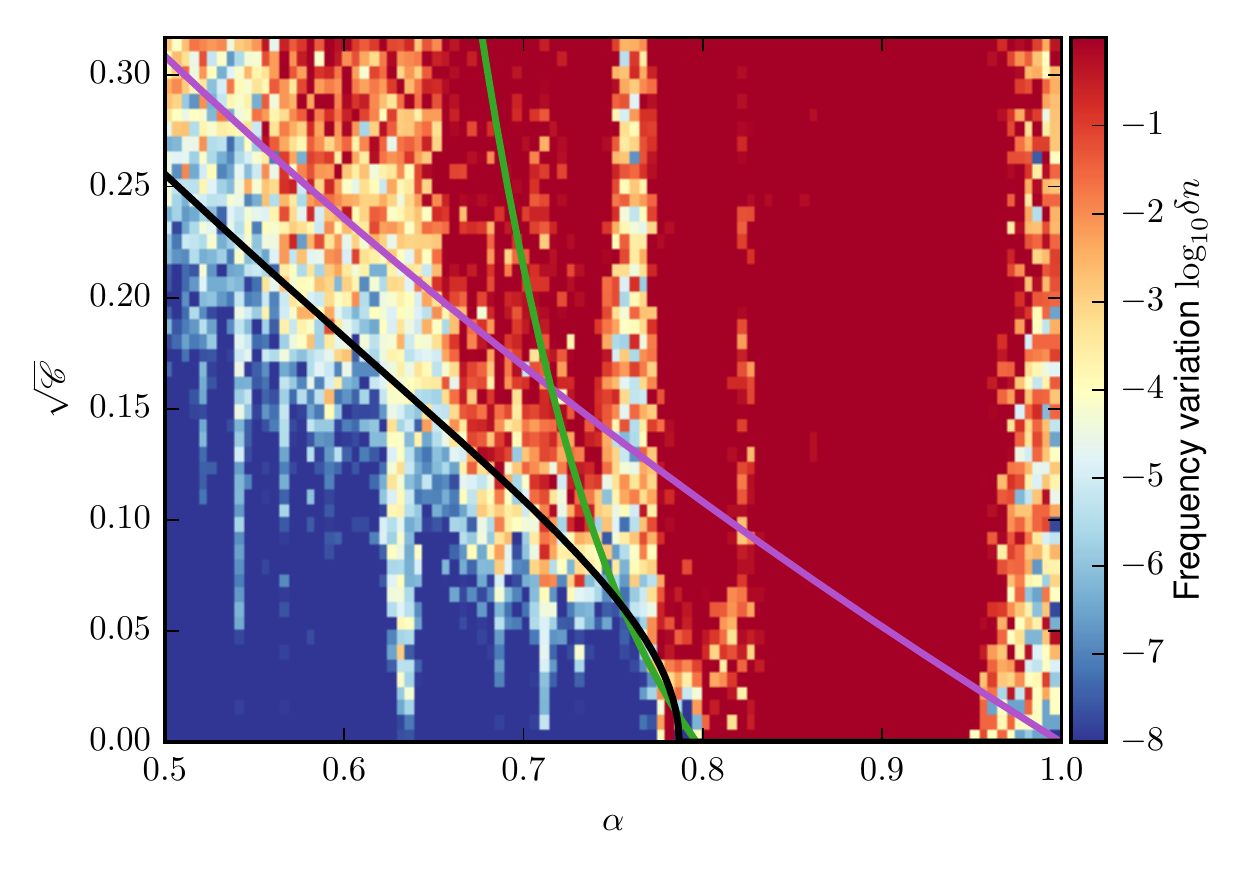}
	\caption{Frequency variation $\delta n$ (\ref{deltan}) for 40,000 initial conditions binned in a $100\times50$ grid with $\epsilon=\dix{-3}$ and $\gamma=1$ .
		The three curves are similar to figure \ref{naf_5_1}.}
	\label{naf_3_1}
\end{figure}

The results of the frequency analysis are plotted in Figure \ref{naf_3_1}. We observe that the $3:2$ MMR is still very stable in comparison to the surrounding regions.
The $2:1$ MMR appears to create an area of moderate chaos in the Hill stable region but is not as marked in the Hill unstable part.
The coorbital resonance also appears more unstable.

\begin{figure}[tbp]
	\includegraphics[width=9.5cm]{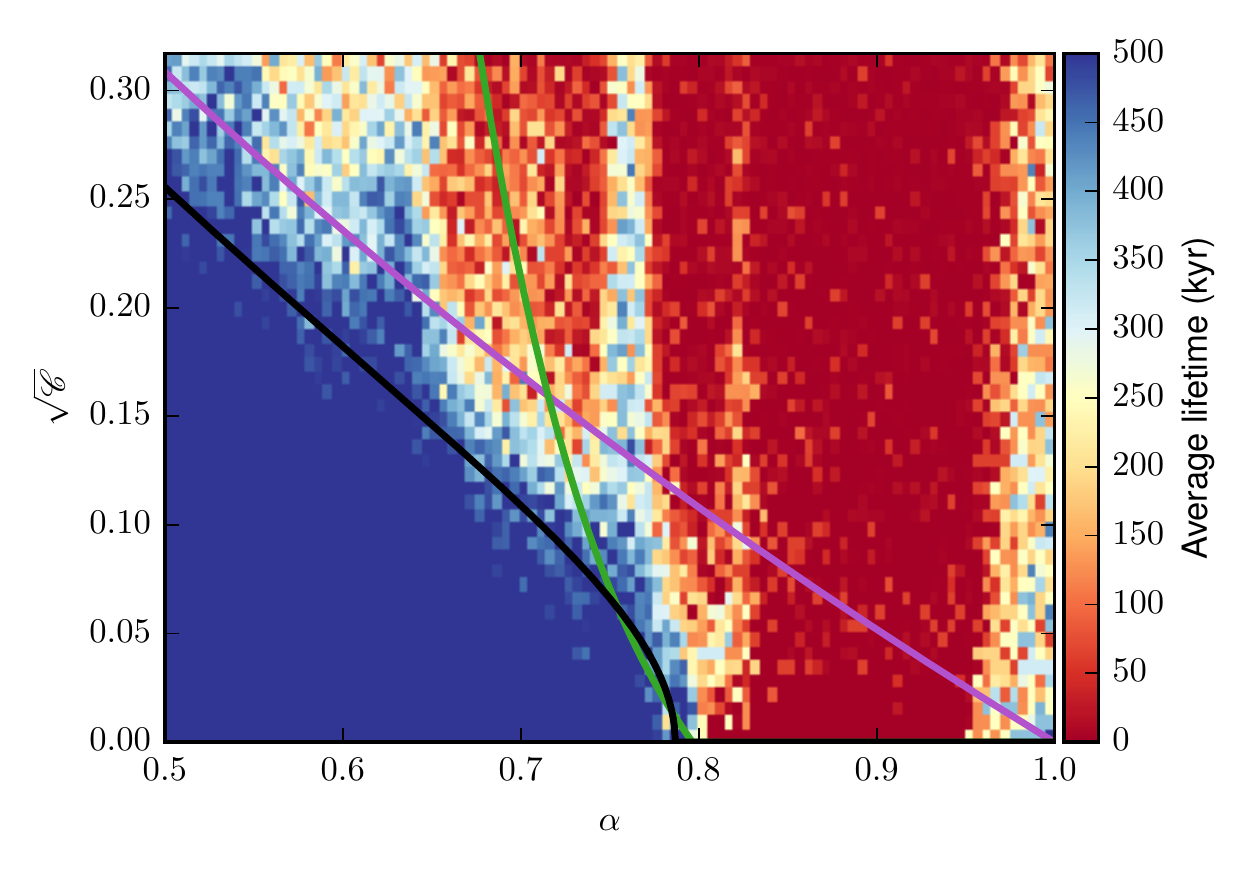}
	\caption{Average lifetime of the system as a function of $\alpha$ and $\rC$. The parameters are similar to figure \ref{naf_3_1}. We see that some Hill stable initial conditions stopped before 500 kyr due to ejections.} 
	\label{crash_3_1}
\end{figure}

We see in Figure \ref{crash_3_1} that some of the Hill stable systems do not survive for 500 kyr. These systems have led to an ejection of the outer planet. As a result, the Hill stability line appears to be more porous in comparison to the case~${\epsilon=\dix{-5}}$.

\begin{figure}[tbp]
	\includegraphics[width=9cm]{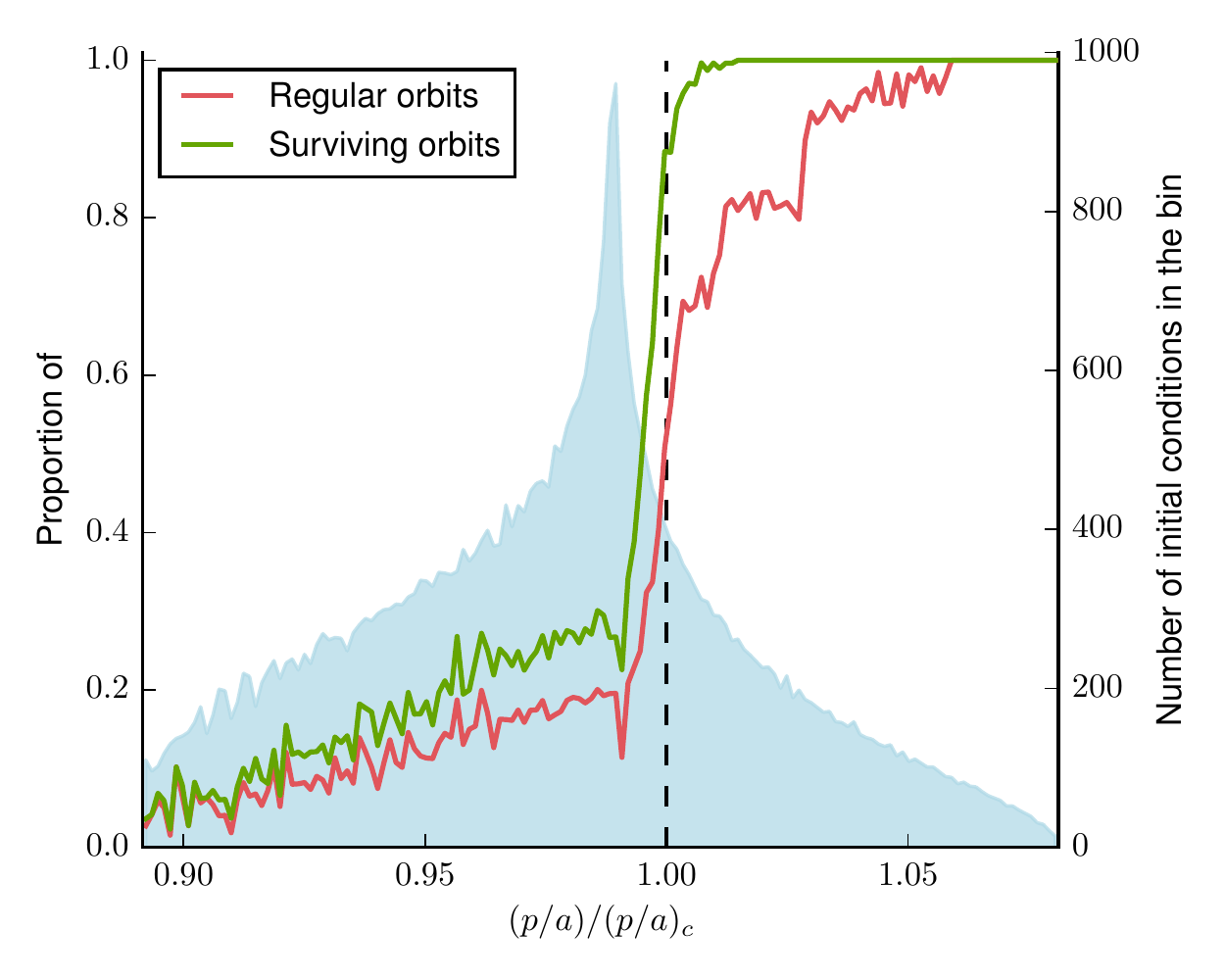}
	\caption{Proportion of orbits that survived for 500 kyr (green curve) and non chaotic orbits ($\delta n <\dix{-4}$, red curve) as a function of $(p/a)/(p/a)_c$.  $(p/a)/(p/a)_c>1$ means that the initial condition verifies the Hill criterion. The light blue histogram represents the number of initial conditions in each of the 150 bins used to compute the fractions plotted.} 
	\label{chaos_3_1}
\end{figure}

We also see that a larger fraction of Hill stable orbits appears to be chaotic. 
This can be quantified thanks to the Figure \ref{chaos_3_1}. We see that the proportion of regular orbits drops before it reaches the Hill stability limit.
However, the more important change of behaviour is still at the Hill stability limit.

\section{Conclusions}

In a two planet system, Hill stability is a topological limit that forbids close encounters between the outer planet and the inner planet or the star.
If verified, the system will remain stable if the outer planet does not escape or if the inner planet does not collide the star.
Moreover, since a minimal distance between planets is imposed, the planet perturbation remains moderate and the system is most likely regular.

We have generalised here Gladman's Hill stability criterion and have shown that it is natural to express the Hill criterion in the AMD framework.
Indeed, we have obtained a simple expression for this criterion (Proposition~\ref{prop.HS}), with only an expansion in $\epsilon$ {(eq. \ref{eps})}. Moreover, it is easy to recover all former published criteria as particular cases of this expression.
Because of its formulation as a function of the AMD, the expression (\ref{hillepsAMD}) is valid in the general spatial case, for any value of eccentricities and inclinations. 
Moreover, our Hill criterion is accurate even for very different planet masses.
We also highlight that the AMD and the semi-major axis ratio $\alpha$ are the main parameters to consider in a stability study.

We show that the Hill stability allows to give an accurate stability limit up to large orbital separations.
The sharp change of behaviour at the Hill stability limit has already been studied in \citep{Barnes2006} or in \citep{Deck2013}.
Nevertheless, our numerical integrations confirm it for a much 
larger range of $\alpha$ and AMD with randomised initial conditions for the parameters not taken into account.

Our simulations for large planet masses  also show that the expansion in $\epsilon$ is valid even for larger planets and that the Hill stability accurately segregates between regular and short-lived initial conditions.
However, it appears that for large mass values, ejections cannot be neglected and a model should be developed to understand this further behaviour.

As shown in several works on tightly packed systems \citep{Chambers1996,Pu2015},
the Gladman's Hill criterion is not adapted anymore in these cases.
Such a sharp limit between almost eternal and short-lived systems no longer exists.
Instead, it appears that there exists a scaling between the initial orbital separation and the time of instability, wider separated orbits becoming unstable after a longer time.
In the cited works, the empirical stability criteria give stability spacing as a function of the mutual radius
\begin{equation}
	\frac{a_{n+1}-a_n}{a_{n+1}}>KR_\text{H},
\end{equation}
where $R_\text{H}=(\epsilon/3)^{1/3}$ is the Hill radius.

In the context of multiplanetary systems, an analytical work on longterm stability is still necessary.
This will be the subject of  future work.

\begin{acknowledgements}
This work was granted access to the HPC resources of MesoPSL financed
by the Region Ile de France and the project Equip@Meso (reference
ANR-10-EQPX-29-01) of the programme Investissements d’Avenir supervised
by the Agence Nationale pour la Recherche. We acknowledge the support of the PNP and of the CS of the Observatoire de Paris \end{acknowledgements}

\bibliographystyle{aa}
\bibliography{Hill}

\appendix
\section{Computation of $R$ at the Lagrange points}
\label{app.R}

The function $R(x,y)=(\rho/\nu)^2$ defined in (\ref{funR}) admits three saddle points situated on the $x$ axis which are the Lagrange points $L_1,L_2$ and $L_3$.
If $x_j$ is the abscissa of the point $L_j$, we have $0<x_1<1$, $x_2>1$ and $x_3<0$.
The $x_j$ quantities only depends on the mass ratio $\epsilon$ and on
\begin{equation}
\zeta =\frac{\gamma}{\gamma+1}= \frac{m_1}{m_1+m_2}.
\end{equation}
 They are the roots of the three different polynomial equations
\begin{align}
(1+\zeta\epsilon)x_1^5-(2+3\zeta\epsilon)x_1^4+\nnb
(1+2\epsilon+\zeta\epsilon)x_1^3-(1+3\epsilon(1-\zeta))x_1^2+\nnb
(2+3\epsilon(1-\zeta))x_1-(1+\epsilon(1-\zeta))&=0,\\
(1+\zeta\epsilon)x_2^5-(2+3\zeta\epsilon)x_2^4+\nnb
(1+3\zeta\epsilon)x_2^3-(1+3\epsilon-\zeta\epsilon))x_2^2+\nnb
(2+3\epsilon(1-\zeta))x_2-(1+\epsilon(1-\zeta))&=0,\\
(1+\zeta\epsilon)x_3^5-(2+3\zeta\epsilon)x_3^4+\nnb
(1+3\zeta\epsilon)x_3^3+(1+3\epsilon(1-\zeta))x_3^2-\nnb
(2+3\epsilon(1-\zeta))x_3+(1+\epsilon(1-\zeta))&=0,
\end{align}
At up to terms of order $\epsilon^{4/3}$, we have
\begin{align}
x_1 &= 1- \left(\frac{\epsilon}{3}\right)^{1/3}+\frac{\gamma+2}{3(\gamma+1)}\left(\frac{\epsilon}{3}\right)^{2/3}+\frac{(\gamma-2)\epsilon}{27(\gamma+1)}+\O(\epsilon^{4/3}),\nnb
x_2 &=  1- \left(\frac{\epsilon}{3}\right)^{1/3}+\frac{\gamma+2}{3(\gamma+1)}\left(\frac{\epsilon}{3}\right)^{2/3}-\frac{(\gamma-2)\epsilon}{27(\gamma+1)}+\O(\epsilon^{4/3}),\label{app.xj}\hspace{-0.5cm}\\
x_3 &= - 1+\frac{7}{12}\frac{\gamma-1}{\gamma+1}\epsilon +\O(\epsilon^2).\nonumber
\end{align}
We can then evaluate $R$ (eq. \ref{funR}) at those points and we have at order, $\epsilon^{4/3}$
\begin{align}
R(L_1)&=1+\frac{3^{4/3}\epsilon^{2/3}\gamma}{(\gamma+1)^2}-\frac{(11+7\gamma)\gamma\epsilon}{3(\gamma+1)^3}+\O(\epsilon^{4/3}),\nnb
R(L_2)&=1+\frac{3^{4/3}\epsilon^{2/3}\gamma}{(\gamma+1)^2}-\frac{(11\gamma+7)\gamma\epsilon}{3(\gamma+1)^3}+\O(\epsilon^{4/3}),\\
R(L_3)&=1+\frac{2\epsilon\gamma}{(\gamma+1)^2} +\O(\epsilon^2)\nonumber.
\end{align}

\section{Expansion of $C_c^\text{Ex}$ and $h_1$}
\label{app.pertterm}

In section \ref{sec.theory}, the Hill stability criterion (\ref{hillepsAMD}) is obtained by the expansion at the leading order in $\epsilon$ of 
\begin{equation}
F=(\gamma+1)^{3/2}\sqrt{\frac{R(L_1)(1+\epsilon\gamma/(\gamma+1)^2)^3}{(1+\epsilon)\left(\gamma/\alpha+1+h_1\right)}}.
\label{app.hillAMD}
\end{equation}
In $F$, the main term depending on $\epsilon$ comes from the expansion of
\begin{equation}
R(L_1)=1+\frac{3^{4/3}\gamma\epsilon^{2/3}}{(\gamma+1)^2}-\frac{(11+7\gamma)}{3(\gamma+1)}\frac{\gamma\epsilon}{(\gamma+1)^2}+\O(\epsilon^{4/3}).
\end{equation}
However, we also need to make sure that $h_1$ remains small in comparison to this term.
Thus, the expansion of $F$ requires an estimate of $h_1$ with respect to $\epsilon$ and $\gamma$.
As explained in section \ref{sec.theory}, $h_1$ is the renormalized perturbation part of the Hamiltonian (\ref{ham})
\begin{equation}
h_1=-\frac{2\Lambda_2^2}{m_2^3\mu^2}\left(\frac{1}{2}\frac{\norm{\tr_1+\tr_2}^2}{m_0}-\frac{\Gr m_1m_2}{\rot}\right).
\label{apph1}
\end{equation}
If we note $\dot{\r}_j=\tr_j/m_j$ and simplify the expression (\ref{apph1}), we obtain $h_1=h_1^\text{T}+h_1^\text{P}$, where
\begin{equation}
h_1^\text{T}=-\frac{\epsilon a_2}{\mu}\frac{\|\gamma \dot{\r}_1+\dot{\r}_2\|^2}{\gamma+1},\ \text{and}\ h_1^\text{P}=\frac{2a_2}{\rot}\frac{\epsilon\gamma}{\gamma+1}.
\label{app.kin}
\end{equation}
As one can see on figure \ref{crithill}, the value of $\epsilon$ is only relevant for small values of the critical AMD $C^{\mathrm{Ex}}_c$, \emph{i.e.}
for close planets.
Using the expansion in $1-\alpha$ (\ref{hillepsalp}), 
we can estimate that we need to compute $h_1$ for systems such that $\rC$ is of order~$\epsilon^{2/3}$.
It corresponds to systems with eccentricities of order $\epsilon^{1/3}$.
We can therefore use the circular approximation in our estimation of $h_1$.
In particular we have $r_j=a_j(1+\O(\epsilon^{1/3}))$.

We first consider the term coming from the gravitational interaction between the two planets, $h_1^\text{P}$.
If the system is Hill stable, the distance $\rot$ is greater than the radius of the Hill sphere $S_\mathrm{H_1}$
\begin{align}
\rot>r_1\max_{j=1,2}|1-x_j|&=r_1(\epsilon/3)^{1/3}+\O(\epsilon^{2/3})\nnb 
&= a_1(\epsilon/3)^{1/3} + \O(\epsilon^{2/3}),
\end{align}
where $x_j$ is defined in (\ref{app.xj}). 
For all times, we therefore have
\begin{equation}
h_1^\text{P}=\frac{2a_2}{\rot}\frac{\epsilon\gamma}{\gamma+1}\leq \frac{2\times 3^{1/3}}{\alpha} \epsilon^{2/3}\frac{\gamma}{\gamma+1} +\O(\epsilon).
\end{equation} 
The gravitational potential term $h_1^\text{P}$ is at most of the same order as the leading term in $\epsilon$ of $R(L_1)$. 
However, we can always choose to estimate the energy and actions values when the two planets are far from each other.
In this case $a_2/\rot=\O(1)$ and $h_1^\text{P}$ is linear in $\epsilon$.
We will from now assume that we have 
\begin{equation}
{h_1^\text{P} = \frac{\epsilon\gamma}{\gamma+1} p_{12}},
\label{app.estP}
\end{equation}
with $p_{12}=\O(1)$.

Moreover, $F$ is a decreasing function of $h_1$, so $C^{\mathrm{Ex}}_c$ is an increasing function of $h_1$. 
Since $h_1^\text{P}$ is positive, neglecting it is equivalent to have a more conservative criterion.

Let us now consider the kinetic term $h_1^\text{T}$.
We develop (\ref{app.kin}) for almost circular orbits. In this limit, we have ${\|\dot{\r}_j\|=\sqrt{\mu/a_j}}+\O(\epsilon^{1/3})$.
We have
\begin{equation}
h_1^\text{T}=-\frac{\epsilon}{\gamma+1}\left(\frac{\gamma^2}{\alpha} +1+\frac{2\gamma}{\sqrt{\alpha}}\cos(\lambda_1-\lambda_2)\right) +\O(\epsilon^{4/3}).
\label{app.devh1}
\end{equation}
Therefore,  $h_1^\text{T}$ is always linear in $\epsilon$. Combining the estimations (\ref{app.estP}) and (\ref{app.devh1}), we see that $h_1=\O(\epsilon)$. 

We can now use the expansions of $h_1$, (\ref{app.devh1}) and (\ref{app.estP}) to obtain the expansion of $F$ in function of $\epsilon$ and $\gamma$.
Because of the term  $3^{4/3}\epsilon^{2/3}\gamma/(\gamma+1)^2$ from $R(L_1)$, we do not keep terms of order ${\O(\epsilon\gamma/(\gamma+1)^2)}$. We keep all other terms depending on $\epsilon$ up to the order $\O(\epsilon^{4/3})$. We obtain

\begin{equation}
F=\sqrt{\frac{\alpha(\gamma+1)^{3}}{\gamma+\alpha}\left(1+\frac{3^{4/3}\epsilon^{2/3}\gamma}{(\gamma+1)^2}\right)D}+\O\left(\frac{\epsilon\gamma}{(\gamma+1)^2},\epsilon^{4/3}\right),
\label{app.Fdev}
\end{equation}
where $D=(1-\epsilon)\left(1-\Frac{\alpha h_1}{\gamma+\alpha}\right)$ and 
\begin{equation*}
\O\left(\frac{\epsilon\gamma}{(\gamma+1)^2},\epsilon^{4/3}\right)=\O\left(\frac{\epsilon\gamma}{(\gamma+1)^2}\right)+\O\left(\epsilon^{4/3}\right).
\end{equation*}
Let us develop $D$ at the same order than $F$. We have
\begin{align*}
D&=1-\epsilon-\frac{\alpha h_1^\text{T}}{\gamma+\alpha}+\O\left(\frac{\epsilon\gamma}{(\gamma+1)^2},\epsilon^{4/3}\right)\nnb
 &=1+\epsilon\left(\frac{\gamma^2+\alpha}{(\gamma+1)(\gamma+\alpha)} -1\right)+\O\left(\frac{\epsilon\gamma}{(\gamma+1)^2},\epsilon^{4/3}\right)\nnb
 \end{align*}
  \begin{align}
D &=1-\frac{\epsilon\gamma}{(\gamma+1)^2}\frac{(\gamma+1)(\alpha+1)}{\gamma+\alpha}+\O\left(\frac{\epsilon\gamma}{(\gamma+1)^2},\epsilon^{4/3}\right)\nnb
 &=1+\O\left(\frac{\epsilon\gamma}{(\gamma+1)^2},\epsilon^{4/3}\right).
\end{align}

We see that $D$ has no contribution to $F$ at the considered orders. Therefore, we can write 
\begin{equation}
F=\sqrt{\frac{\alpha(\gamma+1)^{3}}{\gamma+\alpha}\left(1+3^{4/3}\frac{\epsilon^{2/3}\gamma}{(\gamma+1)^2}\right)}+\O\left(\frac{\epsilon\gamma}{(\gamma+1)^2},\epsilon^{4/3}\right)\hspace{-0.05cm}.\hspace{-0.5cm}
\label{app.FdevnoD}
\end{equation}

As an immediate consequence of (\ref{app.FdevnoD}), the expression proposed in (\ref{hillepsAMD}) remains valid in the case of a very uneven mass distribution between the two planets (\emph{e.g.} for ${\O(\epsilon^{1/3})<\gamma<\O(\epsilon^{-1/3}})$).

\end{document}